\documentclass[aps,prb,twocolumn,groupedaddress,showpacs,floatfix]{revtex4}
\usepackage{epsfig}
\usepackage{epsf} 
\begin{document}
%

\title{Transition pressures and enthalpy barriers for the
  cd$\to${$\beta$}-tin transition in Si and Ge under non-hydrostatic
  conditions} 

\author{Katalin Ga\'al-Nagy}
\altaffiliation{present address: Dipartimento di Fisica, 
Universit\`a degli Studi di Milano, via Celoria 16, I-20133 Milano, Italy}
\email{katalin.gaal-nagy@physik.uni-regensburg.de}
\author{Dieter Strauch}
\affiliation{Institut f{\"u}r theoretische Physik, Universit{\"a}t
  Regensburg, D-93040 Regensburg, Germany}

\date{\today}
%
\begin{abstract}
  We present an {\it ab-initio} study of the phase transition
  cd$\to${$\beta$}-tin in Si and Ge under hydrostatic and non-hydrostatic
  pressure. For this purpose we have developed a new method to
  calculate the influence of non-hydrostatic pressure components not
  only on the transition pressure but also on the enthalpy barriers
  between the phases. We find good agreement with available
  experimental and other theoretical data. The calculations have been
  performed using the plane-wave pseudopotential approach to the
  density-functional theory within the local-density and the
  generalized-gradient approximation implemented in VASP.
\end{abstract}
\maketitle

%
%
\section{Introduction}\label{Intro}
The phase transitions in silicon (Si) and germanium (Ge) from the cd
phase (cubic-diamond structure) to the $\beta$-tin phase
(body-centered tetragonal structure, bct) are two of the most studied
solid-solid phase transitions in condensed matter physics, both,
experimentally\cite{McM93} \cite{McM94} \cite{Hu86} \cite{Vor03}
\cite{Zha86} \cite{Oli84} \cite{Wer82} \cite{Men86} \cite{Yos97}
\cite{Spa84} \cite{Cic03} \cite{Wen62} \cite{Jam63} \cite{Asa78}
\cite{Dyu78} \cite{Gup80} \cite{Bau82} \cite{Men83} \cite{Qua83}
\cite{Hu84} \cite{Ton92} \cite{Heb03} \cite{Voh86} and theoretically
\cite{Yin80} \cite{Yin80b} \cite{Yin82} \cite{Yin82a} \cite{Nee84}
\cite{Bis84} \cite{Cha84} \cite{Cha86} \cite{Voh86} \cite{Bis87}
\cite{Met89b} \cite{Boy91} \cite{Miz94} \cite{Nee95} \cite{Mol95}
\cite{Cor96} \cite{Pfr97} \cite{Lee97b} \cite{Chr99} \cite{Gaa99}
\cite{Ack01} \cite{Heb01} \cite{Gaa01} \cite{Muj03} \cite{Gaa04a}
\cite{Gaa04c} \cite{Kac05}. In the experiment, the phase transition in
Si occurs at around 110~kbar and in Ge at around 105~kbar, where also
lower values of the transition pressure are obtained. These lower
values are often considered as caused by non-hydrostatic conditions,
which are able to reduce the transition pressure.\cite{Jam63}

In fact, the pressure in the anvil cell is not exactly
hydrostatic. Usually at pressures up to 100~kbar the
pressure-transmitting medium yields nearly hydrostatic
conditions.\cite{Pie73} Above 150~kbar a non-hydrostatic pressure
profile is visible, and at very high pressures the
pressure-transmitting medium becomes solid which causes a strong
non-hydrostatic effect. Even in the hydrostatic pressure regime
there is a small pressure gradient.\cite{Bar73} Non-hydrostatic
pressure profiles can also be an effect of the geometry of the
cell.\cite{Bri88} Because of relaxation phenomena which happen in
the pressure-transmitting medium the time for compressing and
decompressing has an influence on the measurement.

In theoretical investigations generally hydrostatic conditions are
assumed. Within calculations using the local-density approximation
(LDA) the calculated transition pressures vary between 70 and
99~kbar for Si and between 73 and 98~kbar for Ge. Usually the
transition pressure is strongly underestimated by LDA calculations
whereas calculations using generalized-gradient approximation (GGA)
match the experimental value better (102--164~kbar for Si and
96--118~kbar for Ge). In any case, the discrepancy between
experimental and theoretical results can also be due to
non-hydrostatic pressure conditions in the experiment. {\it
Ab-initio} calculations considering non-hydrostatic pressure are
rare\cite{Che01,Che03} and deal just with the transition
pressure. The influence of non-hydrostatic conditions on the
enthalpy barrier between the two phases is not studied within an
{\it ab-initio} calculation until now. Therefore, we developed a new
method to calculate the transition pressure as well as the enthalpy
barriers also for non-hydrostatic conditions. In a first step, we
obtain the transition pressure and the enthalpy barrier between both
phases as a function of pressure starting from a complete numerical
equation of state for hydrostatic conditions. Here a complete
equation of state means a continuous, multivalued function $V(p)$
where $V$ is the volume and $p$ the pressure, similar to the one
of the common textbook example of the van der Waals gas. In a second
step, this procedure is extended to non-hydrostatic conditions.

This contribution is organized as follows: Firstly, a short
introduction into the theoretical background of the calculations and
the properties of the unit cell used for our calculations is given
(Section~\ref{method}). Then we explain the procedure of
calculating a complete equation of state from a given energy surface
(Section~\ref{Hydr}). Following this, we present results for the
influence of non-hydrostatic pressure components on the transition
pressure and on the height of the enthalpy barrier
(Section~\ref{Nonhydr}). Finally, after a discussion of the results
and comparison with available theoretical data
(Section~\ref{Discussion}) we describe possible extensions of our
procedure and summarize (Section~\ref{Conclusions}).

%
%
\section{Method}\label{method}
We have carried out calculations with the Vienna {\it ab-initio}
simulation package (VASP).\cite{Kre93b,Kre96a,Kre93a,Kre96b} It is
based one a plane-wave pseudopotential approach to the
density-functional theory (DFT).\cite{Hoh64,Koh65} We have used
ultrasoft Vanderbilt-type pseudopotentials \cite{Van90} as supplied
by Kresse and Hafner \cite{Kre94}. The exchange-correlation
potential has been calculated within the GGA due to Perdew and
Wang\cite{Per92b} for Si and Ge and the LDA\cite{Per81,Cep80} for Si
only. The forces on the atoms are derived from a generalized form
\cite{Goe92,Kre93a} of the Hellmann-Feynman theorem \cite{Fey39}
including Pulay forces.\cite{Pul69} For the ultrasoft
pseudopotentials a kinetic-energy cutoff of 270~eV (410~eV) for Si
(Ge) has been sufficient for convergence of the total energy and
provides an error smaller than 0.5~kbar (0.2~kbar) for Si (Ge) to
the pressure according to the Pulay stress.\cite{Pul69} The
special-point summation required a 18$\times$18$\times$18 (24$\times$24$\times$24) mesh
of Monkhorst-Pack points\cite{Mon76} which amounts to 864 (1962)
{\bf k}-points in the irreducible wedge of the Brillouin zone for Si
(Ge). Since the $\beta$-tin phase is metallic we have used a
Methfessel-Paxton smearing\cite{Met89} with a width of 0.2~eV,
including the cd phase, since it is not a priori clear whether a
given set of volume $V$ and ratio $c/a$ of lattice constants yields
a metallic or a semiconducting phase.

In order to minimize an energy offset between the structures it is
important to describe the structures of both phases within the same
bct cell (lattice constants $a=b\not=c$) with two atoms in the basis
at $(0,0,0)$ and $(0,0.5a,0.25c)$. The symmetry of the cd phase
requires $c/a=\sqrt{2}$, whereas $c/a$ can vary for the $\beta$-tin
phase. Using the bct cell the structure of the cd phase with respect
to the conventional face-centered cubic cell is rotated by $45^{\circ}$
around the $c$-axis. In the following, CD and BCT denote the
structure of the cd- and the $\beta$-tin phase using the bct cell.

%
%

\section{Complete equation of state under hydrostatic
conditions}\label{Hydr}
Neglecting temperature and zero-point motion effects, the phase with
the lowest enthalpy $H=E+pV$ is stable. Therefore, the transition
pressure $p^{\rm t}$ for a first-order pressure-induced phase
transition from the cd phase to the $\beta$-tin phase can be determined
from the crossing of the corresponding enthalpy curves $H^{\rm
cd}(p)$ and $H^{\beta\mbox{-}{\rm tin}}(p)$ with $H^{\rm cd}(p^{\rm
t})=H^{\beta\mbox{-}{\rm tin}}(p^{\rm t})$. First-order phase
transitions are accompanied by a discontinuity in the volume $\Delta V$
and an overcoming of an enthalpy barrier which is located between
the phases and which has a height of $\Delta H$.

Under {\it hydrostatic} conditions the pressure is defined as $p=-\partial
E/\partial V$. It can also be determined from the stress tensor
\mbox{\boldmath $\sigma$}.\cite{Nye69,Cha95} Since the structures here
are orthogonal, the off-diagonal components of \mbox{\boldmath $\sigma$}
vanish, and \mbox{\boldmath $\sigma$} has the form
\begin{eqnarray} \label{GlStress}
  \mbox{\boldmath $\sigma$} = 
  \left( 
    \begin{array}{ccc}
      \sigma_{11} &             &            \\ 
                  & \sigma_{22} &            \\
                  &             & \sigma_{33}
    \end{array}
  \right)
  =
  \left( 
    \begin{array}{ccc}
      -p_x &      &     \\
           & -p_y &     \\
           &      & -p_z 
    \end{array}
  \right) \quad .
\end{eqnarray} 
We are using a tetragonal cell, and thus $p_x=p_y$. Under
hydrostatic conditions all three components are equal and correspond
to the external pressure $p$,
\begin{eqnarray}\label{GlHydrBed}
  p_x=p_y=p_z=p \quad .
\end{eqnarray}
Under {\it non-hydrostatic} conditions the average pressure is
defined as
\begin{eqnarray}\label{GlAverageP}
  p_0 = \frac{1}{3} (p_x + p_y + p_z) 
      = - \frac{1}{3}\, {\rm tr}\, \mbox{\boldmath $\sigma$} \quad,
\end{eqnarray}
which is again equal to the external pressure in the case of
hydrostatic conditions.

\begin{figure}[ht]
  \epsfig{figure=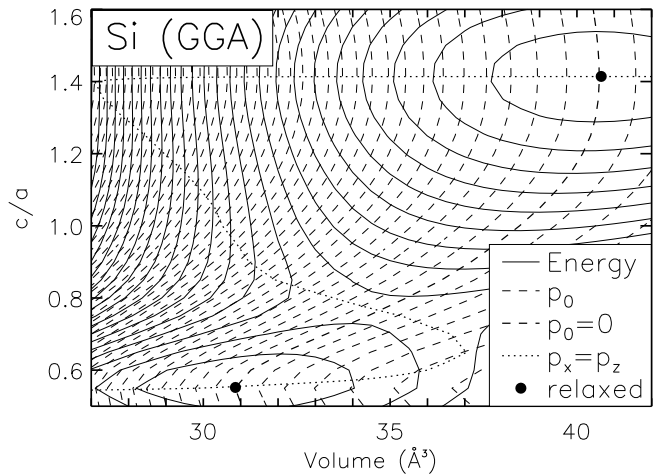,width=8.6cm}
  \caption{Contour plot of the total energy $E(V, c/a)$ (solid
    lines) and of the average pressure $p_0(V,c/a)$ (dashed lines,
    see Eq.~(\ref{GlAverageP})) for Si with GGA. The bold-dashed line
    corresponds to the value $p_0=0$. The interval of the contour
    lines is 50~meV for the energy and 20~kbar for the pressure
    surfaces. The black dots mark the equilibrium positions of the
    cd ($c/a=\sqrt{2}$) and the $\beta$-tin phase ($c/a=0.55$). The
    dotted line marks the parameters under hydrostatic condition.
  }\label{Pic_GefccE}
\end{figure}

We have calculated the total energy as a function of $V$ and
$c/a$. The corresponding energy surface $E(V,c/a)$ is shown in
Fig.~\ref{Pic_GefccE} for Si using the GGA (similar results are
obtained for Ge within GGA and for Si within LDA). The two local
energy minima, according to the two phases, with a saddle in between
are visible. The pressure $p_0(V,c/a)$ is obtained from
Eq.~(\ref{GlStress}) and Eq.~(\ref{GlAverageP}) and is included in
the figure. Except along the (dotted) hydrostatic line with
$p_x=p_z$ the pressure is non-hydrostatic. The local minima are at
the crossing of the $p_0=0$ and the hydrostatic $p_x=p_z$ lines
which defines the equilibrium position. A similar graph can be drawn
also for the enthalpy.\cite{Gaa04c}

\begin{figure}[ht]
  \epsfig{figure=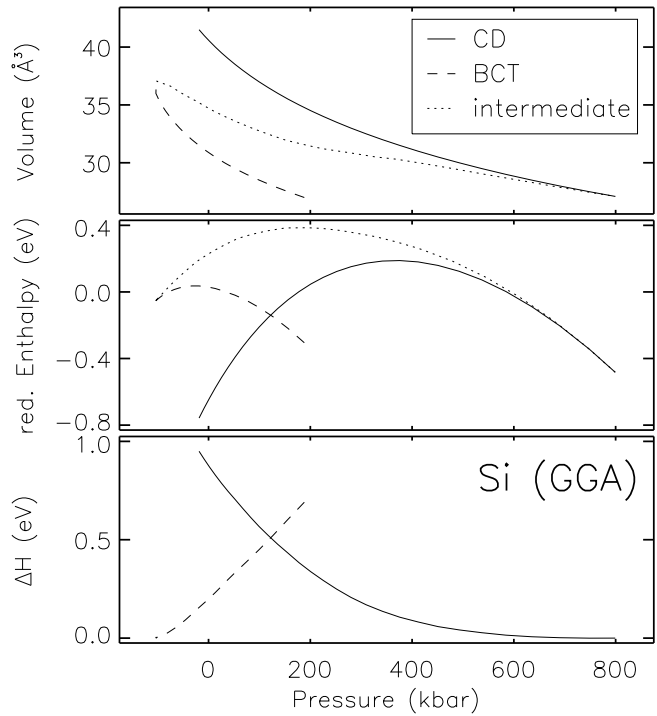, width=8.6cm} 
  \caption{
    Volume $V$, reduced enthalpy (see text), and enthalpy barrier
    $\Delta H$ as a function of the hydrostatic pressure for
    Si within GGA. The crossing of the solid and the dashed line
    determine the transition pressure.
  }\label{Pic_GePVHP} 
\end{figure}

Along the hydrostatic line $p_x=p_z$ the structures are in a local
equilibrium, meaning the total sum of internal and external forces
caused by the pressure is zero. Hence, this condition can be used
to extract the total energy $E$, the external pressure $p_0=p$, and
the volume $V$ from Fig.~\ref{Pic_GefccE} in order to derive a full
equation of state $V(p)$. Also $H(p)$ and, furthermore, the values
along the line across the saddle are accessible. These curves are
shown in Fig.~\ref{Pic_GePVHP}. Here, the ideal cd structure
($c/a=\sqrt{2}$) has been reached only within an error of 1\% for
the lattice parameters. In order to discriminate the enthalpy curves
against each other we have reduced them by a linear background. The
local stability is in accordance with the fact that the $V(p)$
curves are monotonically decreasing and the $H(p)$-curves are
convex. This is in contrast to the the textbook example, the van der
Waals gas, where the line corresponding to the dotted line of $H(p)$
is concave and signals local instability.

The transition pressure $p^{\rm t}$ obtained from the crossing of
the enthalpy curves are listed in Table~\ref{TabHydr}. The
corresponding change $\Delta V$ in the volume at the phase transition can
be read from the upper panel of Fig.~\ref{Pic_GePVHP} as the
difference between $V^{\rm cd}(p^{\rm t})$ (solid line) and
$V^{\beta{\rm -tin}}(p^{\rm t})$ (dashed line). Analogously the enthalpy
barrier $\Delta H$ can be determined from the figures. In order to check
the reliability of this method we compare the results with our
previous ones \cite{Gaa04c} based on the same total-energy
calculations but obtained with a different method to evaluate the
transitions pressures and enthalpy barriers. The agreement is very
good, and the small differences are due to numerical errors. Thus,
we can trust in the new method developed here.

\begin{table}[h]
  \caption{Transition pressures $p^{\rm t}$, volume changes $\Delta V$,
    and enthalpy barriers $\Delta H$ derived from the complete equation
    of state in comparison with our previous results obtained with
    an alternative method (in
    parenthesis).\cite{Gaa04c}}\label{TabHydr} 
  \begin{ruledtabular}
    \begin{tabular}{l|c|c|c}
                          &Ge-GGA &Si-GGA & Si-LDA\\
      \hline
      $p^{\rm t}$ (kbar)  &  96  (96) & 122 (121) &  80  (79) \\
      $\Delta V$ (\AA$^3$)       &  7.5 (7.5)& 8.3 (8.3) &  8.5 (8.5)\\
      $\Delta H$ (meV)         &  421 (423)& 510 (515) &  502 (508)\\
    \end{tabular}
  \end{ruledtabular}
\end{table}

Since we have determined a complete equation of state, we can
calculate also the enthalpy barrier as a function of pressure. We
have to distinguish between the barrier for the cd$\to${$\beta$}-tin
transition, approaching from the cd phase, and the one for the
{$\beta$}-tin$\to$cd transition, approaching from the $\beta$-tin phase. In
general, the enthalpy barrier $\Delta H$ has its origin in the energy
saddle between the two phases and can be calculated as the
difference of the reduced enthalpy of the phases (solid and dashed
lines, respectively) and the one from the saddle (dotted line), see
Fig.~\ref{Pic_GePVHP}. In particular, the enthalpy barrier from the
cd phase is the difference between the solid and the dotted line,
and the one from the $\beta$-tin phase is the difference between the
dashed and the dotted line. At the transition pressure $p^{\rm t}$
the enthalpy barriers from both phases have the same height. The
determination of the enthalpy barrier as a function of pressure is
important to estimate the barrier in the case of over- and
underpressurizing the medium. Hence, the phase transitions will
happen at a pressure different from $p^{\rm t}$ which results in a
different height of the barrier. As expected, the enthalpy barrier
from the cd phase is decreasing with increasing pressure whereas the
one from the $\beta$-tin phase decreases with decreasing pressure. At
zero pressure there is still an enthalpy barrier left for the
$\beta$-tin$\to$cd transition. This points at the fact that there is no
spontaneous transition $\beta$-tin$\to$cd. In the experiment the phase
transition cd$\to${$\beta$}-tin is irreversible.

%
%

\section{Phase transition under non-hydrostatic
    conditions}\label{Nonhydr} 
\begin{figure}[ht]
  \epsfig{figure=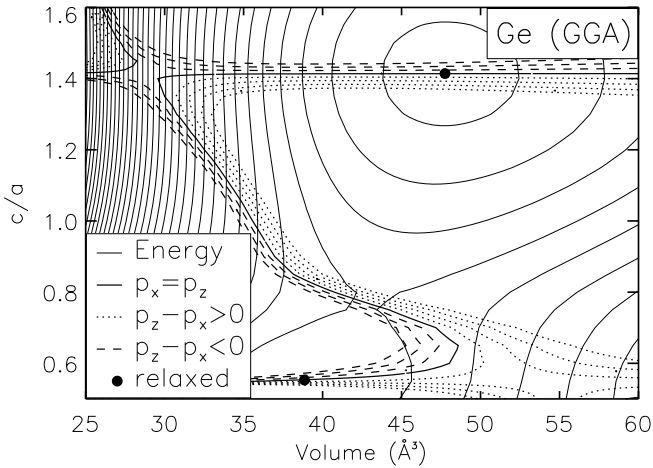, width=8.6cm}
  \caption{Contour plot of the total energy $E(V, c/a)$ (solid
    lines) for Ge (GGA) with an interval of the contour lines of
    0.2~eV. Besides the hydrostatic condition (bold solid line)
    non-hydrostatic conditions ($p_z-p_x=-15, -10, \ldots ,20$) are shown.
  } \label{Pic_GefccNHE}
\end{figure}

The procedure for determining transition pressures and enthalpy
barriers described in the previous section can be extended to the
case of non-hydrostatic pressure. Besides the hydrostatic condition
$p_z-p_x=0$ the values for non-hydrostatic pressure components
$p_z-p_x=d\not=0$ (with a fixed value of $d$) can be extracted from
the energy surface $E(V,c/a)$ along the corresponding lines of
Fig.~\ref{Pic_GefccNHE}.  A first naive trial considering just the
total energy $E^{\rm nh}$ under non-hydrostatic conditions and the
corresponding values $p^{\rm nh}_0$ for the average pressure and
$V^{\rm nh}$ for the volume gives wrong results, e.g., an increase
of the transition pressure for $p_z>p_x$ and $p_z<p_x$. This is in
contrast to the experimental observations. Thus is is necessary to
include strain effects.

Similar to the stress tensor of Eq.~(\ref{GlStress}) the strain
tensor \mbox{\boldmath $\epsilon$} can be reduced to a diagonal form for
orthogonal systems\cite{Nye69,Lan70,Cha95}
\begin{eqnarray} 
  \mbox{\boldmath $\epsilon$} = 
  \left( 
    \begin{array}{ccc}
      \epsilon_{11} &             &            \\ 
                  & \epsilon_{22} &            \\
                  &             & \epsilon_{33}
    \end{array}
  \right)
  =
  \left( 
    \begin{array}{ccc} 
      \epsilon_x &            &     \\
                 & \epsilon_y &     \\
                 &            & \epsilon_z 
    \end{array}
  \right) 
\end{eqnarray} 
where $\epsilon_x$, $\epsilon_y$, and $\epsilon_z$ are along the cartesian crystal axes.
For small stress and homogeneous strain the components of
\mbox{\boldmath $\epsilon$} can be derived as\cite{Lan70,Nye69}
\begin{eqnarray} \label{EqHomSm}
  \epsilon_{jj}= \frac{x_j' - {x_j}}{x_j} \quad ,
\end{eqnarray}
where $x_j$ is the lattice parameter in the $j$-direction. Here
$x_j$ corresponds to the unstrained and $x_j'$ to the strained
crystal. Including stress and strain the enthalpy can be written
as\cite{Lan70}
\begin{eqnarray} \label{GlHStrain}
  \tilde{H} = \tilde{E} 
  + \sum_{j=1}^{3}\sigma_{jj}\epsilon_{jj} \quad , 
\end{eqnarray}
where $\tilde{H}$ is the enthalpy and $\tilde{E}$ the total energy
per volume. The calculation of the enthalpy at non-hydrostatic
stress is based on Eq.~(\ref{GlHStrain}). The numerical realization
is described in Appendix~\ref{StrainCalculation}.

\begin{figure}[ht]
  \epsfig{figure=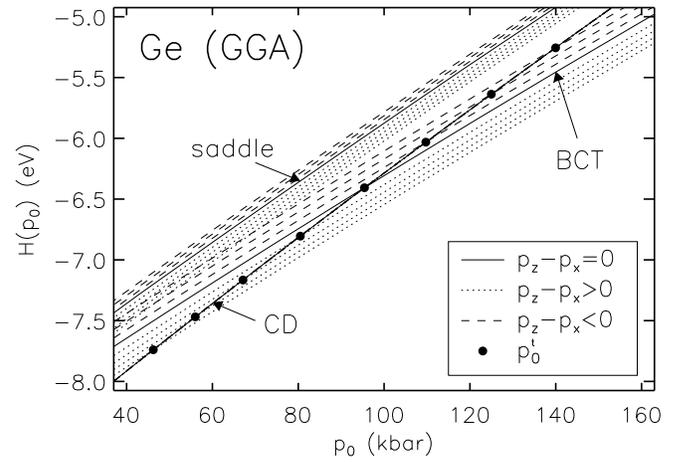, width=8.6cm}
  \caption{Equation of state $H(p_0)$ for non-hydrostatic conditions
    as a function of the average pressure $p_0$. The difference
    $p_z-p_x$ of two neighboring lines is 5~kbar. The black dots
    mark the transition pressures $p_0^{\rm t}$.
  }\label{Pic_GeStrainHP}
\end{figure}

It turns out that the strain-only contribution to the enthalpy
vanishes for the cd phase, that it is linear with the pressure for
the $\beta$-tin phase, and that it is non-linear for the contribution
along the line across the saddle. This effect is apparent in
Fig.~\ref{Pic_GeStrainHP}, where the enthalpy including strain is
presented. Since there is no strain effect on the cd phase, the
change of the transition pressure is due to the strained $\beta$-tin
phase. From Fig.~\ref{Pic_GeStrainHP} we can find the transition
pressures for fixed non-hydrostatic conditions in the same manner
as mentioned in the previous section. 

\begin{figure}[h!]
  \epsfig{figure=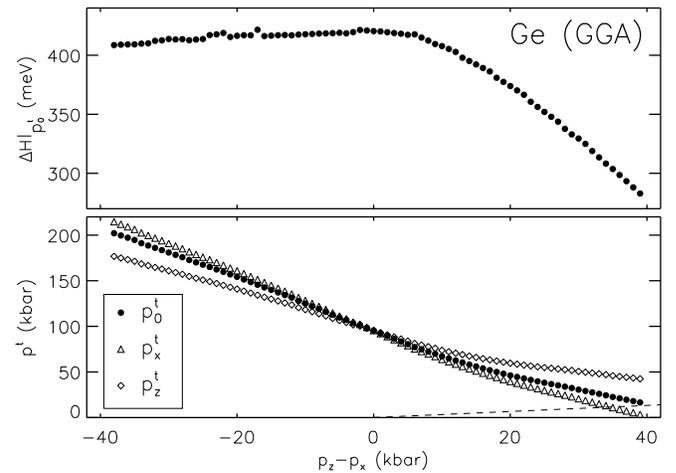, width=8.6cm}
  \caption{Enthalpy barriers at the average transition pressures
    $p^{\rm t}_0$ and transition pressures (average transition
    pressure $p^{\rm t}_0$ and the corresponding components 
    $p^{\rm t}_x$ and $p^{\rm t}_z$) 
   as a function of $p_z-p_x$. The dashed line marks
    the boundary $p_x=0$ and $p_z=0$.
  }\label{Pic_GeStrainBarrPt}
\end{figure}

In addition to the average transition pressure $p_0^{\rm t}$ also
their components $p^{\rm t}_x$ and $p^{\rm t}_z$ in the $x$ and $z$
directions are shown as well as the enthalpy barrier at the phase
transition as a function of $p_z-p_x$ in
Fig.~\ref{Pic_GeStrainBarrPt}. For the transition pressure one has
the relation $p_0^{\rm t}=(2p^{\rm t}_x+p^{\rm t}_z)/3$. The
boundary for the lowest pressure is fixed by the condition that the
components of $p_0$ are not negative, since we exclude a stretching
of the crystal. We find a strong lowering of the transition pressure
if the pressure in the $z$ direction is larger than in the $x$ and
$y$ directions. Thus, the crystal is more stable against compression
along the $x$- and $y$-axes in the case of a strong non-hydrostatic
component in these directions which cause an increase of the
transition pressure. The corresponding enthalpy barriers are
lowering in any case, but their value remains still larger than the
thermal energy at room temperature (RT).

Besides the non-hydrostatic effects we can consider finally also the
case of over- and under-pressurization of the crystal. To this end
calculations of the enthalpy barriers as a function of the average
pressure and non-hydrostatic conditions have been carried out.  At
very high pressures and very large non-hydrostatic components the
enthalpy barrier for the cd$\to${$\beta$}-tin transition is smaller than
the thermal energy at RT, but these conditions do not appear by
chance in the experiment; instead, they have to be applied by
intention. In contrast, the enthalpy barrier is never smaller than
25~meV for the $\beta$-tin$\to$cd transition even at the largest
non-hydrostatic components with $p_x$, $p_y$, and $p_z$ not negative
(no stretching).
\begin{figure}[t]
  \epsfig{figure=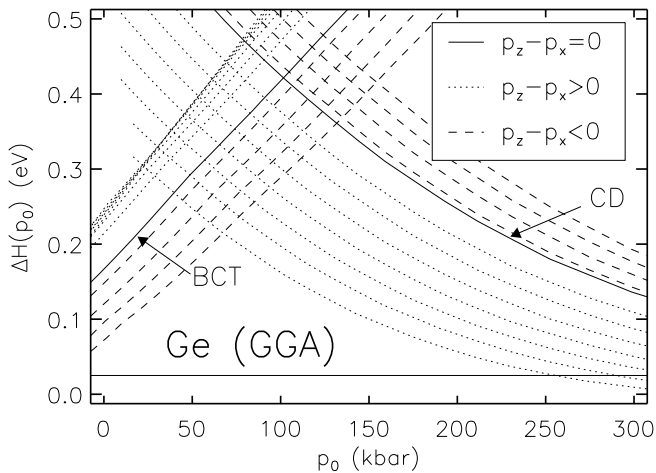,width=8.6cm}
  \caption{Enthalpy barriers like in Fig.~\ref{Pic_GePVHP} for fixed
    non-hydrostatic conditions $p_z-p_x=-20, -15, \ldots
  35$~kbar. The solid horizontal line at 25~meV marks the thermal
  excitation energy at RT. 
  }\label{Pic_GeGGAStrainBarrP}
\end{figure}

%
%
\section{Discussion of the results}\label{Discussion}
In the past, non-hydrostatic conditions have been investigated for
different reasons.\cite{Lee97,Che01,Che03,Wan93,Wan95,Lib00}
Directly comparable with our results are just the ones from Lee {\it
et al.}\ \cite{Lee97} and Cheng {\it et al.}\ \cite{Che01,Che03}
Besides the transition pressures also the function $p^{\rm
t}_z(p^{\rm t}_x)$ is given in these contributions. This function
can be obtained from our results by a linear fit of the values for
the transition pressure in Fig.~\ref{Pic_GeStrainBarrPt}. Already
Lee {\it et al.}\ \cite{Lee97} found a linear relation of $p^{\rm
t}_z$ and $p^{\rm t}_x$. In their molecular-dynamics investigation
they found $p^{\rm t}=127$~kbar for hydrostatic conditions and
$p_z^{\rm t}=p_x^{\rm t} + 90$~kbar. The additive constant
corresponds to the lowest possible transition pressure. On the
contrary, Cheng {\it et al.}\ \cite{Che01,Che03} obtained $p^{\rm
t}=95$~kbar and $p_z^{\rm t}=0.737~p_x^{\rm t} + 25$~kbar for Ge,
and $p^{\rm t}=114$~kbar and $p_z^{\rm t}=0.658~p_x^{\rm t} +
39$~kbar for Si. Although the last values have been obtained also
with VASP using GGA, the results for the transition pressure are
slightly different from our results (95.9~kbar for Ge(GGA),
122.1~kbar for Si(GGA), and 79.6~kbar for Si(LDA)). This can be due
to the fact that they have used different cells for the phases and
also different pseudopotentials and convergence parameters. The
choice of different unit cells can lead to an energy offset between
the energy curves to which the transition pressure is very
sensitive. Nevertheless, our results for the linear functions which
are $p_z^{\rm t}=0.651~p_x^{\rm t} + 35.1$~kbar for Ge(GGA),
$p_z^{\rm t}=0.606~p_x^{\rm t} + 47.3$~kbar for Si(GGA), and
$p_z^{\rm t}=0.619~p_x^{\rm t} + 29.8$~kbar for Si(LDA),
respectively, agree very well with the ones of
Refs.~\onlinecite{Che01} and \onlinecite{Che03}. The difference of
the additive constants rely on the different values of the
transition pressures in the hydrostatic case. Since Cheng {\it et
al.}\ \cite{Che01,Che03} restricted themselves to the enthalpy
difference between the phases using path integrals, the enthalpy
barrier was not accessible to them.

The experimental values for the transition pressures vary between
103 and 133~kbar for Si\cite{McM93, McM94, Hu86, Oli84, Wer82,
Zha86, Vor03} and between 103 and 110~kbar for Ge\cite{Oli84, Men86,
Yos97, Wer82, Spa84, Zha86,Cic03} where the firmed values are at
around 110~kbar and 105~kbar, respectively. In both cases, our
results obtained with GGA agree perfectly, whereas the LDA result
underestimates the experimental value, which is a well known
problem.

The good agreement of our results with the ones of Cheng {\it et
al.}\ \cite{Che01,Che03} confirm the reliability of our method,
which provides a larger field of applications. In addition, our
method can be extended to, e.g., the $\beta$-tin$\to${\it Imma}$\to$sh
transitions in Si and Ge. After the extraction of a two-dimensional
energy surface from a three-dimensional one using the values along
the lines where two components of the stress tensor are equal (like
in our previous work\cite{Gaa04c,Gaa04d}), the method mentioned here
can be applied to this extracted surface. By the choice of two equal
components the main pressure direction is chosen. Further extensions
even to non-orthorhombic structures are possible, too.

%
%
\section{Summary}\label{Conclusions}
We have developed a new method for investigating first-order
high-pressure phase transitions which is based on the calculation of
a complete equation of state. Besides the transition pressure and
the volume change, which are also available with the common-tangent
construction, the enthalpy barrier between the phases can be
obtained with our method. A comparison with results for Si and Ge
from common methods shows the reliability of the new method. Further
on, the enthalpy barrier can be determined as a function of the
external pressure which makes effects from over- and
underpressurizing accessible. An extension of this method allows
us also to investigate high-pressure phase transitions under
non-hydrostatic conditions, in particular the transition pressure
and the enthalpy barrier, which are both decreasing if the pressure
component along the $c$-axis is larger than the other ones.  Our
results show an excellent agreement with available experimental and
theoretical data. This new method can be extended also to other
phase transitions and also to ones including orthorhombic
structures, for example, the transitions $\beta$-tin$\to${\it
Imma}$\to$sh. Thus, we have developed a powerful tool for
investigating phase transitions under hydrostatic and
non-hydrostatic conditions.

%
%

\section*{Acknowledgment}
Support by the Heinrich B\"oll Stiftung, Germany, is gratefully
acknowledged. This work was funded in part by the EU's 6th Framework
Programme through the NANOQUANTA Network of Excellence
(NMP4-CT-2004-500198).

%
%
\begin{appendix}
%
%
\section{Calculation of the enthalpy including
  stress}\label{StrainCalculation} 
Here we give a short description of the formulae used for the
calculation of the enthalpy including stress and strain effects
starting from Eq.~(\ref{GlHStrain}).
\parbox{7.5cm}{\begin{eqnarray*}
  {\rm d}H 
  &=& {\rm d}E \\
  &+& \left(p_x \frac{{\rm d}a}{a_0}  
    +p_y\frac{{\rm d}b}{b_0} 
    +p_z \frac{{\rm d}c}{c_0}\right) V_0 \quad ,
\end{eqnarray*}}\hfill
\parbox{8mm}{ \begin{eqnarray} \end{eqnarray}} \\
which is equivalent to ${\rm d}H={\rm d}E+p\,{\rm d}V$ in the case of
hydrostatic conditions. Since Eq.~(\ref{EqHomSm}) holds just for
small stress, we need to integrate this equation and cannot go
directly to the absolute values. The integration is performed using
the recursively defined equation
\parbox{7.5cm}{\begin{eqnarray*}
  H^{i} &=& (E^{i}_{\rm nh}-E^{i-1}_{\rm nh}) +
  V^{i-1}(p_0^{i}-p_0^{i-1}) \\ 
  &+& V^{i-1}
    \sum_{j=x,y,z}
    \frac{x_{j}^{i}-x_{j}^{i-1}}{x_{j}^{i-1}}p_{j}^{i} \\
  &+&   H^{i-1}
\end{eqnarray*}}\hfill
\parbox{8mm}{ \begin{eqnarray}\label{GlRec} \end{eqnarray} } \\
where $x_{j}$ are the lattice constants along the three cartesian
directions $x,y$ and $z$, and the difference from the previous step
$(i-1)$ is calculated along a line $p_x-p_z=d$ for fixed
non-hydrostatic conditions (see Fig.~\ref{Pic_GefccNHE}) starting
from the equilibrium structure of the cd phase. $E_{\rm nh}$ is here
the total energy along a line $p_x-p_z=d$. The enthalpy $H(p)$ under
non-hydrostatic conditions corresponds to the calculated points
$H^i(p^i)$. By symmetrizing this equation numerical errors have been
reduced.

\end{appendix}
%
%

\bibliography{bib}

\begin{thebibliography}{78}
\expandafter\ifx\csname natexlab\endcsname\relax\def\natexlab#1{#1}\fi
\expandafter\ifx\csname bibnamefont\endcsname\relax
  \def\bibnamefont#1{#1}\fi
\expandafter\ifx\csname bibfnamefont\endcsname\relax
  \def\bibfnamefont#1{#1}\fi
\expandafter\ifx\csname citenamefont\endcsname\relax
  \def\citenamefont#1{#1}\fi
\expandafter\ifx\csname url\endcsname\relax
  \def\url#1{\texttt{#1}}\fi
\expandafter\ifx\csname urlprefix\endcsname\relax\def\urlprefix{URL }\fi
\providecommand{\bibinfo}[2]{#2}
\providecommand{\eprint}[2][]{\url{#2}}

\bibitem[{\citenamefont{McMahon and Nelmes}(1993)}]{McM93}
\bibinfo{author}{\bibfnamefont{M.~I.} \bibnamefont{McMahon}} \bibnamefont{and}
  \bibinfo{author}{\bibfnamefont{R.~J.} \bibnamefont{Nelmes}},
  \bibinfo{journal}{Phys.~Rev.~B} \textbf{\bibinfo{volume}{47}},
  \bibinfo{pages}{R8337} (\bibinfo{year}{1993}).

\bibitem[{\citenamefont{McMahon et~al.}(1994)\citenamefont{McMahon, Nelmes,
  Wright, and Allan}}]{McM94}
\bibinfo{author}{\bibfnamefont{M.~I.} \bibnamefont{McMahon}},
  \bibinfo{author}{\bibfnamefont{R.~J.} \bibnamefont{Nelmes}},
  \bibinfo{author}{\bibfnamefont{N.~G.} \bibnamefont{Wright}},
  \bibnamefont{and} \bibinfo{author}{\bibfnamefont{D.~R.} \bibnamefont{Allan}},
  \bibinfo{journal}{Phys.~Rev.~B} \textbf{\bibinfo{volume}{50}},
  \bibinfo{pages}{739} (\bibinfo{year}{1994}).

\bibitem[{\citenamefont{Hu et~al.}(1986)\citenamefont{Hu, Merkle, Menoni, and
  Spain}}]{Hu86}
\bibinfo{author}{\bibfnamefont{J.~Z.} \bibnamefont{Hu}},
  \bibinfo{author}{\bibfnamefont{L.~D.} \bibnamefont{Merkle}},
  \bibinfo{author}{\bibfnamefont{C.~S.} \bibnamefont{Menoni}},
  \bibnamefont{and} \bibinfo{author}{\bibfnamefont{I.~L.} \bibnamefont{Spain}},
  \bibinfo{journal}{Phys.~Rev.~B} \textbf{\bibinfo{volume}{34}},
  \bibinfo{pages}{4679} (\bibinfo{year}{1986}).

\bibitem[{\citenamefont{Voronin et~al.}(2003)\citenamefont{Voronin, Pantea,
  Zerda, Wang, and Zhao}}]{Vor03}
\bibinfo{author}{\bibfnamefont{G.~A.} \bibnamefont{Voronin}},
  \bibinfo{author}{\bibfnamefont{C.}~\bibnamefont{Pantea}},
  \bibinfo{author}{\bibfnamefont{T.~W.} \bibnamefont{Zerda}},
  \bibinfo{author}{\bibfnamefont{L.}~\bibnamefont{Wang}}, \bibnamefont{and}
  \bibinfo{author}{\bibfnamefont{Y.}~\bibnamefont{Zhao}},
  \bibinfo{journal}{Phys.~Rev.~B} \textbf{\bibinfo{volume}{68}},
  \bibinfo{pages}{020102(R)} (\bibinfo{year}{2003}).

\bibitem[{\citenamefont{Zhao et~al.}(1986)\citenamefont{Zhao, Buehler, Sites,
  and Spain}}]{Zha86}
\bibinfo{author}{\bibfnamefont{Y.-X.} \bibnamefont{Zhao}},
  \bibinfo{author}{\bibfnamefont{F.}~\bibnamefont{Buehler}},
  \bibinfo{author}{\bibfnamefont{J.~R.} \bibnamefont{Sites}}, \bibnamefont{and}
  \bibinfo{author}{\bibfnamefont{I.~L.} \bibnamefont{Spain}},
  \bibinfo{journal}{Solid~State~Commun.~} \textbf{\bibinfo{volume}{59}},
  \bibinfo{pages}{679} (\bibinfo{year}{1986}).

\bibitem[{\citenamefont{Olijnyk et~al.}(1984)\citenamefont{Olijnyk, Sikka, and
  Holzapfel}}]{Oli84}
\bibinfo{author}{\bibfnamefont{H.}~\bibnamefont{Olijnyk}},
  \bibinfo{author}{\bibfnamefont{S.~K.} \bibnamefont{Sikka}}, \bibnamefont{and}
  \bibinfo{author}{\bibfnamefont{W.~B.} \bibnamefont{Holzapfel}},
  \bibinfo{journal}{Phys.~Lett.~} \textbf{\bibinfo{volume}{103 A}},
  \bibinfo{pages}{137} (\bibinfo{year}{1984}).

\bibitem[{\citenamefont{Werner et~al.}(1982)\citenamefont{Werner, Sanjurjo, and
  Cardona}}]{Wer82}
\bibinfo{author}{\bibfnamefont{A.}~\bibnamefont{Werner}},
  \bibinfo{author}{\bibfnamefont{J.~A.} \bibnamefont{Sanjurjo}},
  \bibnamefont{and} \bibinfo{author}{\bibfnamefont{M.}~\bibnamefont{Cardona}},
  \bibinfo{journal}{Solid~State~Commun.~} \textbf{\bibinfo{volume}{44}},
  \bibinfo{pages}{155} (\bibinfo{year}{1982}).

\bibitem[{\citenamefont{Menoni et~al.}(1986)\citenamefont{Menoni, Hu, and
  Spain}}]{Men86}
\bibinfo{author}{\bibfnamefont{C.~S.} \bibnamefont{Menoni}},
  \bibinfo{author}{\bibfnamefont{J.~Z.} \bibnamefont{Hu}}, \bibnamefont{and}
  \bibinfo{author}{\bibfnamefont{I.~L.} \bibnamefont{Spain}},
  \bibinfo{journal}{Phys.~Rev.~B} \textbf{\bibinfo{volume}{34}},
  \bibinfo{pages}{362} (\bibinfo{year}{1986}).

\bibitem[{\citenamefont{Yoshiasa et~al.}(1997)\citenamefont{Yoshiasa, Koto,
  Maeda, , and Ishii}}]{Yos97}
\bibinfo{author}{\bibfnamefont{A.}~\bibnamefont{Yoshiasa}},
  \bibinfo{author}{\bibfnamefont{K.}~\bibnamefont{Koto}},
  \bibinfo{author}{\bibfnamefont{H.}~\bibnamefont{Maeda}}, , \bibnamefont{and}
  \bibinfo{author}{\bibfnamefont{T.}~\bibnamefont{Ishii}},
  \bibinfo{journal}{Jpn.~J.~Appl.~Phys.~} \textbf{\bibinfo{volume}{36}},
  \bibinfo{pages}{781} (\bibinfo{year}{1997}).

\bibitem[{\citenamefont{Spain et~al.}(1984)\citenamefont{Spain, Hu, Menoni, and
  Black}}]{Spa84}
\bibinfo{author}{\bibfnamefont{I.~L.} \bibnamefont{Spain}},
  \bibinfo{author}{\bibfnamefont{J.~Z.} \bibnamefont{Hu}},
  \bibinfo{author}{\bibfnamefont{C.~S.} \bibnamefont{Menoni}},
  \bibnamefont{and} \bibinfo{author}{\bibfnamefont{D.}~\bibnamefont{Black}},
  \bibinfo{journal}{J.~Phys. (Paris)} \textbf{\bibinfo{volume}{45}},
  \bibinfo{pages}{Suppl.~Colloq.~C8, C8} (\bibinfo{year}{1984}).

\bibitem[{\citenamefont{Cicco et~al.}(2003)\citenamefont{Cicco, Frasini,
  Minicucci, Principi, Iti{\'e}, and Munsch}}]{Cic03}
\bibinfo{author}{\bibfnamefont{A.~D.} \bibnamefont{Cicco}},
  \bibinfo{author}{\bibfnamefont{A.~C.} \bibnamefont{Frasini}},
  \bibinfo{author}{\bibfnamefont{M.}~\bibnamefont{Minicucci}},
  \bibinfo{author}{\bibfnamefont{E.}~\bibnamefont{Principi}},
  \bibinfo{author}{\bibfnamefont{J.-P.} \bibnamefont{Iti{\'e}}},
  \bibnamefont{and} \bibinfo{author}{\bibfnamefont{P.}~\bibnamefont{Munsch}},
  \bibinfo{journal}{Phys.~Stat.~Sol.~(b)} \textbf{\bibinfo{volume}{240}},
  \bibinfo{pages}{19} (\bibinfo{year}{2003}).

\bibitem[{\citenamefont{Wentorf and Kasper}(1963)}]{Wen62}
\bibinfo{author}{\bibfnamefont{R.~H.} \bibnamefont{Wentorf}} \bibnamefont{and}
  \bibinfo{author}{\bibfnamefont{J.~S.} \bibnamefont{Kasper}},
  \bibinfo{journal}{Science} \textbf{\bibinfo{volume}{139}},
  \bibinfo{pages}{338} (\bibinfo{year}{1963}).

\bibitem[{\citenamefont{Jamieson}(1963)}]{Jam63}
\bibinfo{author}{\bibfnamefont{J.~C.} \bibnamefont{Jamieson}},
  \bibinfo{journal}{Science} \textbf{\bibinfo{volume}{139}},
  \bibinfo{pages}{762} (\bibinfo{year}{1963}).

\bibitem[{\citenamefont{Asaumi and Minomura}(1978)}]{Asa78}
\bibinfo{author}{\bibfnamefont{K.}~\bibnamefont{Asaumi}} \bibnamefont{and}
  \bibinfo{author}{\bibfnamefont{S.}~\bibnamefont{Minomura}},
  \bibinfo{journal}{J.~Phys.~Soc.~Jpn.~} \textbf{\bibinfo{volume}{45}},
  \bibinfo{pages}{1061} (\bibinfo{year}{1978}).

\bibitem[{\citenamefont{Dyuzheva et~al.}(1978)\citenamefont{Dyuzheva,
  Kabalkina, and Novichkov}}]{Dyu78}
\bibinfo{author}{\bibfnamefont{T.~I.} \bibnamefont{Dyuzheva}},
  \bibinfo{author}{\bibfnamefont{S.~S.} \bibnamefont{Kabalkina}},
  \bibnamefont{and} \bibinfo{author}{\bibfnamefont{V.~P.}
  \bibnamefont{Novichkov}}, \bibinfo{journal}{Sov.~Phys.~JETP}
  \textbf{\bibinfo{volume}{47}}, \bibinfo{pages}{931} (\bibinfo{year}{1978}).

\bibitem[{\citenamefont{Gupta and Ruoff}(1980)}]{Gup80}
\bibinfo{author}{\bibfnamefont{M.~C.} \bibnamefont{Gupta}} \bibnamefont{and}
  \bibinfo{author}{\bibfnamefont{A.~L.} \bibnamefont{Ruoff}},
  \bibinfo{journal}{J.~Appl.~Phys.~} \textbf{\bibinfo{volume}{51}},
  \bibinfo{pages}{1072} (\bibinfo{year}{1980}).

\bibitem[{\citenamefont{Baublitz and Ruoff}(1982)}]{Bau82}
\bibinfo{author}{\bibfnamefont{M.~A.} \bibnamefont{Baublitz}} \bibnamefont{and}
  \bibinfo{author}{\bibfnamefont{A.~L.} \bibnamefont{Ruoff}},
  \bibinfo{journal}{J.~Appl.~Phys.~} \textbf{\bibinfo{volume}{53}},
  \bibinfo{pages}{5669} (\bibinfo{year}{1982}).

\bibitem[{\citenamefont{Menoni et~al.}(1983)\citenamefont{Menoni, Hu, and
  Spain}}]{Men83}
\bibinfo{author}{\bibfnamefont{C.~S.} \bibnamefont{Menoni}},
  \bibinfo{author}{\bibfnamefont{J.~Z.} \bibnamefont{Hu}}, \bibnamefont{and}
  \bibinfo{author}{\bibfnamefont{I.~L.} \bibnamefont{Spain}}, in
  \emph{\bibinfo{booktitle}{High Pressure in Science and Technology}}, edited
  by \bibinfo{editor}{\bibfnamefont{C.}~\bibnamefont{Homan}},
  \bibinfo{editor}{\bibfnamefont{R.~K.} \bibnamefont{MacCrone}},
  \bibnamefont{and} \bibinfo{editor}{\bibfnamefont{E.}~\bibnamefont{Walley}}
  (\bibinfo{publisher}{North Holland, Amsterdam}, \bibinfo{year}{1983}).

\bibitem[{\citenamefont{Quadri et~al.}(1983)\citenamefont{Quadri, Skelton, and
  Webb}}]{Qua83}
\bibinfo{author}{\bibfnamefont{S.~B.} \bibnamefont{Quadri}},
  \bibinfo{author}{\bibfnamefont{E.~F.} \bibnamefont{Skelton}},
  \bibnamefont{and} \bibinfo{author}{\bibfnamefont{A.~W.} \bibnamefont{Webb}},
  \bibinfo{journal}{J.~Appl.~Phys.~} \textbf{\bibinfo{volume}{54}},
  \bibinfo{pages}{3609} (\bibinfo{year}{1983}).

\bibitem[{\citenamefont{Hu and Spain}(1984)}]{Hu84}
\bibinfo{author}{\bibfnamefont{J.~Z.} \bibnamefont{Hu}} \bibnamefont{and}
  \bibinfo{author}{\bibfnamefont{I.~L.} \bibnamefont{Spain}},
  \bibinfo{journal}{Solid~State~Commun.~} \textbf{\bibinfo{volume}{51}},
  \bibinfo{pages}{263} (\bibinfo{year}{1984}).

\bibitem[{\citenamefont{Tonkov}(1992)}]{Ton92}
\bibinfo{author}{\bibfnamefont{E.~Y.} \bibnamefont{Tonkov}},
  \emph{\bibinfo{title}{High Pressure Phase Transformations}},
  vol.~\bibinfo{volume}{2} (\bibinfo{publisher}{Gordon and Breach Science
  Publishers, Philadelphia}, \bibinfo{year}{1992}).

\bibitem[{\citenamefont{Hebbache and Zemzemi}(2003)}]{Heb03}
\bibinfo{author}{\bibfnamefont{M.}~\bibnamefont{Hebbache}} \bibnamefont{and}
  \bibinfo{author}{\bibfnamefont{M.}~\bibnamefont{Zemzemi}},
  \bibinfo{journal}{Phys.~Rev.~B} \textbf{\bibinfo{volume}{67}},
  \bibinfo{pages}{233302} (\bibinfo{year}{2003}).

\bibitem[{\citenamefont{Vohra et~al.}(1986)\citenamefont{Vohra, Brister,
  Desgreniers, Ruoff, Chang, and Cohen}}]{Voh86}
\bibinfo{author}{\bibfnamefont{Y.~K.} \bibnamefont{Vohra}},
  \bibinfo{author}{\bibfnamefont{K.~E.} \bibnamefont{Brister}},
  \bibinfo{author}{\bibfnamefont{S.}~\bibnamefont{Desgreniers}},
  \bibinfo{author}{\bibfnamefont{A.~L.} \bibnamefont{Ruoff}},
  \bibinfo{author}{\bibfnamefont{K.~J.} \bibnamefont{Chang}}, \bibnamefont{and}
  \bibinfo{author}{\bibfnamefont{M.~L.} \bibnamefont{Cohen}},
  \bibinfo{journal}{Phys.~Rev.~Lett.~} \textbf{\bibinfo{volume}{56}},
  \bibinfo{pages}{1944} (\bibinfo{year}{1986}).

\bibitem[{\citenamefont{Yin and Cohen}(1980{\natexlab{a}})}]{Yin80}
\bibinfo{author}{\bibfnamefont{M.~T.} \bibnamefont{Yin}} \bibnamefont{and}
  \bibinfo{author}{\bibfnamefont{M.~L.} \bibnamefont{Cohen}},
  \bibinfo{journal}{Phys.~Rev.~Lett.~} \textbf{\bibinfo{volume}{45}},
  \bibinfo{pages}{1004} (\bibinfo{year}{1980}{\natexlab{a}}).

\bibitem[{\citenamefont{Yin and Cohen}(1980{\natexlab{b}})}]{Yin80b}
\bibinfo{author}{\bibfnamefont{M.~T.} \bibnamefont{Yin}} \bibnamefont{and}
  \bibinfo{author}{\bibfnamefont{M.~L.} \bibnamefont{Cohen}},
  \bibinfo{journal}{Solid~State~Commun.~} \textbf{\bibinfo{volume}{38}},
  \bibinfo{pages}{625} (\bibinfo{year}{1980}{\natexlab{b}}).

\bibitem[{\citenamefont{Yin and Cohen}(1982{\natexlab{a}})}]{Yin82}
\bibinfo{author}{\bibfnamefont{M.~T.} \bibnamefont{Yin}} \bibnamefont{and}
  \bibinfo{author}{\bibfnamefont{M.~L.} \bibnamefont{Cohen}},
  \bibinfo{journal}{Phys.~Rev.~B} \textbf{\bibinfo{volume}{26}},
  \bibinfo{pages}{5668} (\bibinfo{year}{1982}{\natexlab{a}}).

\bibitem[{\citenamefont{Yin and Cohen}(1982{\natexlab{b}})}]{Yin82a}
\bibinfo{author}{\bibfnamefont{M.~T.} \bibnamefont{Yin}} \bibnamefont{and}
  \bibinfo{author}{\bibfnamefont{M.~L.} \bibnamefont{Cohen}},
  \bibinfo{journal}{Phys.~Rev.~B} \textbf{\bibinfo{volume}{26}},
  \bibinfo{pages}{3259} (\bibinfo{year}{1982}{\natexlab{b}}).

\bibitem[{\citenamefont{Needs and Martin}(1984)}]{Nee84}
\bibinfo{author}{\bibfnamefont{R.~J.} \bibnamefont{Needs}} \bibnamefont{and}
  \bibinfo{author}{\bibfnamefont{R.~M.} \bibnamefont{Martin}},
  \bibinfo{journal}{Phys.~Rev.~B} \textbf{\bibinfo{volume}{30}},
  \bibinfo{pages}{R5390} (\bibinfo{year}{1984}).

\bibitem[{\citenamefont{Chang and Cohen}(1984)}]{Cha84}
\bibinfo{author}{\bibfnamefont{K.~J.} \bibnamefont{Chang}} \bibnamefont{and}
  \bibinfo{author}{\bibfnamefont{M.~L.} \bibnamefont{Cohen}},
  \bibinfo{journal}{Phys.~Rev.~B} \textbf{\bibinfo{volume}{30}},
  \bibinfo{pages}{R5376} (\bibinfo{year}{1984}).

\bibitem[{\citenamefont{Chang and Cohen}(1986)}]{Cha86}
\bibinfo{author}{\bibfnamefont{K.~J.} \bibnamefont{Chang}} \bibnamefont{and}
  \bibinfo{author}{\bibfnamefont{M.~L.} \bibnamefont{Cohen}},
  \bibinfo{journal}{Phys.~Rev.~B} \textbf{\bibinfo{volume}{34}},
  \bibinfo{pages}{8581} (\bibinfo{year}{1986}).

\bibitem[{\citenamefont{Methfessel et~al.}(1989)\citenamefont{Methfessel,
  Rodriguez, and Andersen}}]{Met89b}
\bibinfo{author}{\bibfnamefont{M.}~\bibnamefont{Methfessel}},
  \bibinfo{author}{\bibfnamefont{C.~O.} \bibnamefont{Rodriguez}},
  \bibnamefont{and} \bibinfo{author}{\bibfnamefont{O.~K.}
  \bibnamefont{Andersen}}, \bibinfo{journal}{Phys.~Rev.~B}
  \textbf{\bibinfo{volume}{40}}, \bibinfo{pages}{R2009} (\bibinfo{year}{1989}).

\bibitem[{\citenamefont{Boyer et~al.}(1991)\citenamefont{Boyer, Kaxiras,
  Feldman, Broughton, and Mehl}}]{Boy91}
\bibinfo{author}{\bibfnamefont{L.~L.} \bibnamefont{Boyer}},
  \bibinfo{author}{\bibfnamefont{E.}~\bibnamefont{Kaxiras}},
  \bibinfo{author}{\bibfnamefont{J.~L.} \bibnamefont{Feldman}},
  \bibinfo{author}{\bibfnamefont{J.~Q.} \bibnamefont{Broughton}},
  \bibnamefont{and} \bibinfo{author}{\bibfnamefont{M.~J.} \bibnamefont{Mehl}},
  \bibinfo{journal}{Phys.~Rev.~Lett.~} \textbf{\bibinfo{volume}{67}},
  \bibinfo{pages}{715} (\bibinfo{year}{1991}).

\bibitem[{\citenamefont{Mizushima et~al.}(1994)\citenamefont{Mizushima, Yip,
  and Kaxiras}}]{Miz94}
\bibinfo{author}{\bibfnamefont{K.}~\bibnamefont{Mizushima}},
  \bibinfo{author}{\bibfnamefont{S.}~\bibnamefont{Yip}}, \bibnamefont{and}
  \bibinfo{author}{\bibfnamefont{E.}~\bibnamefont{Kaxiras}},
  \bibinfo{journal}{Phys.~Rev.~B} \textbf{\bibinfo{volume}{50}},
  \bibinfo{pages}{14952} (\bibinfo{year}{1994}).

\bibitem[{\citenamefont{Needs and Mujica}(1995)}]{Nee95}
\bibinfo{author}{\bibfnamefont{R.~J.} \bibnamefont{Needs}} \bibnamefont{and}
  \bibinfo{author}{\bibfnamefont{A.}~\bibnamefont{Mujica}},
  \bibinfo{journal}{Phys.~Rev.~B} \textbf{\bibinfo{volume}{51}},
  \bibinfo{pages}{9652} (\bibinfo{year}{1995}).

\bibitem[{\citenamefont{Moll et~al.}(1995)\citenamefont{Moll, Bockstedte,
  Fuchs, Pehlke, and Scheffler}}]{Mol95}
\bibinfo{author}{\bibfnamefont{N.}~\bibnamefont{Moll}},
  \bibinfo{author}{\bibfnamefont{M.}~\bibnamefont{Bockstedte}},
  \bibinfo{author}{\bibfnamefont{M.}~\bibnamefont{Fuchs}},
  \bibinfo{author}{\bibfnamefont{E.}~\bibnamefont{Pehlke}}, \bibnamefont{and}
  \bibinfo{author}{\bibfnamefont{M.}~\bibnamefont{Scheffler}},
  \bibinfo{journal}{Phys.~Rev.~B} \textbf{\bibinfo{volume}{52}},
  \bibinfo{pages}{2550} (\bibinfo{year}{1995}).

\bibitem[{\citenamefont{DalCorso et~al.}(1996)\citenamefont{DalCorso,
  Pasquarello, Baldereschi, and Car}}]{Cor96}
\bibinfo{author}{\bibfnamefont{A.}~\bibnamefont{DalCorso}},
  \bibinfo{author}{\bibfnamefont{A.}~\bibnamefont{Pasquarello}},
  \bibinfo{author}{\bibfnamefont{A.}~\bibnamefont{Baldereschi}},
  \bibnamefont{and} \bibinfo{author}{\bibfnamefont{R.}~\bibnamefont{Car}},
  \bibinfo{journal}{Phys.~Rev.~B} \textbf{\bibinfo{volume}{53}},
  \bibinfo{pages}{1180} (\bibinfo{year}{1996}).

\bibitem[{\citenamefont{Pfrommer et~al.}(1997)\citenamefont{Pfrommer, C\^ot\'e,
  Louie, and Cohen}}]{Pfr97}
\bibinfo{author}{\bibfnamefont{B.~G.} \bibnamefont{Pfrommer}},
  \bibinfo{author}{\bibfnamefont{M.}~\bibnamefont{C\^ot\'e}},
  \bibinfo{author}{\bibfnamefont{S.~G.} \bibnamefont{Louie}}, \bibnamefont{and}
  \bibinfo{author}{\bibfnamefont{M.~L.} \bibnamefont{Cohen}},
  \bibinfo{journal}{Phys.~Rev.~B} \textbf{\bibinfo{volume}{56}},
  \bibinfo{pages}{6662} (\bibinfo{year}{1997}).

\bibitem[{\citenamefont{Lee and Martin}(1997)}]{Lee97b}
\bibinfo{author}{\bibfnamefont{I.-H.} \bibnamefont{Lee}} \bibnamefont{and}
  \bibinfo{author}{\bibfnamefont{R.~M.} \bibnamefont{Martin}},
  \bibinfo{journal}{Phys.~Rev.~B} \textbf{\bibinfo{volume}{56}},
  \bibinfo{pages}{7197} (\bibinfo{year}{1997}).

\bibitem[{\citenamefont{Christensen et~al.}(1999)\citenamefont{Christensen,
  Novikov, Alonso, and Rodriguez}}]{Chr99}
\bibinfo{author}{\bibfnamefont{N.~E.} \bibnamefont{Christensen}},
  \bibinfo{author}{\bibfnamefont{D.~L.} \bibnamefont{Novikov}},
  \bibinfo{author}{\bibfnamefont{R.~E.} \bibnamefont{Alonso}},
  \bibnamefont{and} \bibinfo{author}{\bibfnamefont{C.~O.}
  \bibnamefont{Rodriguez}}, \bibinfo{journal}{Phys.~Stat.~Sol.~(b)}
  \textbf{\bibinfo{volume}{211}}, \bibinfo{pages}{5} (\bibinfo{year}{1999}).

\bibitem[{\citenamefont{Ga\'al-Nagy et~al.}(1999)\citenamefont{Ga\'al-Nagy,
  Bauer, Schmitt, Karch, Pavone, and Strauch}}]{Gaa99}
\bibinfo{author}{\bibfnamefont{K.}~\bibnamefont{Ga\'al-Nagy}},
  \bibinfo{author}{\bibfnamefont{A.}~\bibnamefont{Bauer}},
  \bibinfo{author}{\bibfnamefont{M.}~\bibnamefont{Schmitt}},
  \bibinfo{author}{\bibfnamefont{K.}~\bibnamefont{Karch}},
  \bibinfo{author}{\bibfnamefont{P.}~\bibnamefont{Pavone}}, \bibnamefont{and}
  \bibinfo{author}{\bibfnamefont{D.}~\bibnamefont{Strauch}},
  \bibinfo{journal}{Phys.~Stat.~Sol.~(b)} \textbf{\bibinfo{volume}{211}},
  \bibinfo{pages}{275} (\bibinfo{year}{1999}).

\bibitem[{\citenamefont{Ackland}(2001)}]{Ack01}
\bibinfo{author}{\bibfnamefont{G.~J.} \bibnamefont{Ackland}},
  \bibinfo{journal}{Rep.~Prog.~Phys.~} \textbf{\bibinfo{volume}{64}},
  \bibinfo{pages}{483} (\bibinfo{year}{2001}).

\bibitem[{\citenamefont{Hebbache et~al.}(2001)\citenamefont{Hebbache,
  Mattesini, and Szeftel}}]{Heb01}
\bibinfo{author}{\bibfnamefont{M.}~\bibnamefont{Hebbache}},
  \bibinfo{author}{\bibfnamefont{M.}~\bibnamefont{Mattesini}},
  \bibnamefont{and} \bibinfo{author}{\bibfnamefont{J.}~\bibnamefont{Szeftel}},
  \bibinfo{journal}{Phys.~Rev.~B} \textbf{\bibinfo{volume}{63}},
  \bibinfo{pages}{205201} (\bibinfo{year}{2001}).

\bibitem[{\citenamefont{Ga\'al-Nagy et~al.}(2001)\citenamefont{Ga\'al-Nagy,
  Schmitt, Pavone, and Strauch}}]{Gaa01}
\bibinfo{author}{\bibfnamefont{K.}~\bibnamefont{Ga\'al-Nagy}},
  \bibinfo{author}{\bibfnamefont{M.}~\bibnamefont{Schmitt}},
  \bibinfo{author}{\bibfnamefont{P.}~\bibnamefont{Pavone}}, \bibnamefont{and}
  \bibinfo{author}{\bibfnamefont{D.}~\bibnamefont{Strauch}},
  \bibinfo{journal}{Comp.~Mat.~Sci.~} \textbf{\bibinfo{volume}{22}},
  \bibinfo{pages}{49} (\bibinfo{year}{2001}).

\bibitem[{\citenamefont{Mujica et~al.}(2003)\citenamefont{Mujica, Rubio,
  Mu$\tilde{\textsc n}$os, and Needs}}]{Muj03}
\bibinfo{author}{\bibfnamefont{A.}~\bibnamefont{Mujica}},
  \bibinfo{author}{\bibfnamefont{A.}~\bibnamefont{Rubio}},
  \bibinfo{author}{\bibfnamefont{A.}~\bibnamefont{Mu$\tilde{\textsc n}$os}},
  \bibnamefont{and} \bibinfo{author}{\bibfnamefont{R.~J.} \bibnamefont{Needs}},
  \bibinfo{journal}{Rev.~Mod.~Phys.~} \textbf{\bibinfo{volume}{75}},
  \bibinfo{pages}{863} (\bibinfo{year}{2003}).

\bibitem[{\citenamefont{Ga\'al-Nagy
  et~al.}(2004{\natexlab{a}})\citenamefont{Ga\'al-Nagy, Bauer, Pavone, and
  Strauch}}]{Gaa04a}
\bibinfo{author}{\bibfnamefont{K.}~\bibnamefont{Ga\'al-Nagy}},
  \bibinfo{author}{\bibfnamefont{A.}~\bibnamefont{Bauer}},
  \bibinfo{author}{\bibfnamefont{P.}~\bibnamefont{Pavone}}, \bibnamefont{and}
  \bibinfo{author}{\bibfnamefont{D.}~\bibnamefont{Strauch}},
  \bibinfo{journal}{Comp.~Mat.~Sci.~} \textbf{\bibinfo{volume}{30}},
  \bibinfo{pages}{1} (\bibinfo{year}{2004}{\natexlab{a}}).

\bibitem[{\citenamefont{Ga\'al-Nagy
  et~al.}(2004{\natexlab{b}})\citenamefont{Ga\'al-Nagy, Pavone, and
  Strauch}}]{Gaa04c}
\bibinfo{author}{\bibfnamefont{K.}~\bibnamefont{Ga\'al-Nagy}},
  \bibinfo{author}{\bibfnamefont{P.}~\bibnamefont{Pavone}}, \bibnamefont{and}
  \bibinfo{author}{\bibfnamefont{D.}~\bibnamefont{Strauch}},
  \bibinfo{journal}{Phys.~Rev.~B} \textbf{\bibinfo{volume}{69}},
  \bibinfo{pages}{134112} (\bibinfo{year}{2004}{\natexlab{b}}).

\bibitem[{\citenamefont{Kaczmarski et~al.}(2005)\citenamefont{Kaczmarski,
  Bedoya-Mart{\'i}nez, and Hern{\'a}ndez}}]{Kac05}
\bibinfo{author}{\bibfnamefont{M.}~\bibnamefont{Kaczmarski}},
  \bibinfo{author}{\bibfnamefont{O.~N.} \bibnamefont{Bedoya-Mart{\'i}nez}},
  \bibnamefont{and} \bibinfo{author}{\bibfnamefont{E.~R.}
  \bibnamefont{Hern{\'a}ndez}}, \bibinfo{journal}{Phys.~Rev.~Lett.~}
  \textbf{\bibinfo{volume}{94}}, \bibinfo{pages}{095701}
  (\bibinfo{year}{2005}).

\bibitem[{\citenamefont{Biswas et~al.}(1984)\citenamefont{Biswas, Martin,
  Needs, and Nielsen}}]{Bis84}
\bibinfo{author}{\bibfnamefont{R.}~\bibnamefont{Biswas}},
  \bibinfo{author}{\bibfnamefont{R.~M.} \bibnamefont{Martin}},
  \bibinfo{author}{\bibfnamefont{R.~J.} \bibnamefont{Needs}}, \bibnamefont{and}
  \bibinfo{author}{\bibfnamefont{O.~H.} \bibnamefont{Nielsen}},
  \bibinfo{journal}{Phys.~Rev.~B} \textbf{\bibinfo{volume}{30}},
  \bibinfo{pages}{3210} (\bibinfo{year}{1984}).

\bibitem[{\citenamefont{Biswas et~al.}(1987)\citenamefont{Biswas, Martin,
  Needs, and Nielsen}}]{Bis87}
\bibinfo{author}{\bibfnamefont{R.}~\bibnamefont{Biswas}},
  \bibinfo{author}{\bibfnamefont{R.~M.} \bibnamefont{Martin}},
  \bibinfo{author}{\bibfnamefont{R.~J.} \bibnamefont{Needs}}, \bibnamefont{and}
  \bibinfo{author}{\bibfnamefont{O.~H.} \bibnamefont{Nielsen}},
  \bibinfo{journal}{Phys.~Rev.~B} \textbf{\bibinfo{volume}{35}},
  \bibinfo{pages}{9559} (\bibinfo{year}{1987}).

\bibitem[{\citenamefont{Piermarini et~al.}(1973)\citenamefont{Piermarini,
  Block, and Barnett}}]{Pie73}
\bibinfo{author}{\bibfnamefont{G.~J.} \bibnamefont{Piermarini}},
  \bibinfo{author}{\bibfnamefont{S.}~\bibnamefont{Block}}, \bibnamefont{and}
  \bibinfo{author}{\bibfnamefont{J.~S.} \bibnamefont{Barnett}},
  \bibinfo{journal}{J.~Appl.~Phys.~} \textbf{\bibinfo{volume}{44}},
  \bibinfo{pages}{5377} (\bibinfo{year}{1973}).

\bibitem[{\citenamefont{Barnett et~al.}(1973)\citenamefont{Barnett, Block, and
  Piermarini}}]{Bar73}
\bibinfo{author}{\bibfnamefont{J.~S.} \bibnamefont{Barnett}},
  \bibinfo{author}{\bibfnamefont{S.}~\bibnamefont{Block}}, \bibnamefont{and}
  \bibinfo{author}{\bibfnamefont{G.~J.} \bibnamefont{Piermarini}},
  \bibinfo{journal}{Rev.~Sci.~Instrum.~} \textbf{\bibinfo{volume}{44}},
  \bibinfo{pages}{1} (\bibinfo{year}{1973}).

\bibitem[{\citenamefont{Brister et~al.}(1988)\citenamefont{Brister, Vohra, and
  Ruoff}}]{Bri88}
\bibinfo{author}{\bibfnamefont{K.~E.} \bibnamefont{Brister}},
  \bibinfo{author}{\bibfnamefont{Y.~K.} \bibnamefont{Vohra}}, \bibnamefont{and}
  \bibinfo{author}{\bibfnamefont{A.~L.} \bibnamefont{Ruoff}},
  \bibinfo{journal}{Rev.~Sci.~Instrum.~} \textbf{\bibinfo{volume}{59}},
  \bibinfo{pages}{318} (\bibinfo{year}{1988}).

\bibitem[{\citenamefont{Cheng et~al.}(2001)\citenamefont{Cheng, Huang, and
  Li}}]{Che01}
\bibinfo{author}{\bibfnamefont{C.}~\bibnamefont{Cheng}},
  \bibinfo{author}{\bibfnamefont{W.~H.} \bibnamefont{Huang}}, \bibnamefont{and}
  \bibinfo{author}{\bibfnamefont{H.~J.} \bibnamefont{Li}},
  \bibinfo{journal}{Phys.~Rev.~B} \textbf{\bibinfo{volume}{63}},
  \bibinfo{pages}{153202} (\bibinfo{year}{2001}).

\bibitem[{\citenamefont{Cheng}(2003)}]{Che03}
\bibinfo{author}{\bibfnamefont{C.}~\bibnamefont{Cheng}},
  \bibinfo{journal}{Phys.~Rev.~B} \textbf{\bibinfo{volume}{67}},
  \bibinfo{pages}{134109} (\bibinfo{year}{2003}).

\bibitem[{\citenamefont{Kresse and Hafner}(1993)}]{Kre93b}
\bibinfo{author}{\bibfnamefont{G.}~\bibnamefont{Kresse}} \bibnamefont{and}
  \bibinfo{author}{\bibfnamefont{J.}~\bibnamefont{Hafner}},
  \bibinfo{journal}{Phys.~Rev.~B} \textbf{\bibinfo{volume}{47}},
  \bibinfo{pages}{R558} (\bibinfo{year}{1993}).

\bibitem[{\citenamefont{Kresse and Furthm\"uller}(1996{\natexlab{a}})}]{Kre96a}
\bibinfo{author}{\bibfnamefont{G.}~\bibnamefont{Kresse}} \bibnamefont{and}
  \bibinfo{author}{\bibfnamefont{J.}~\bibnamefont{Furthm\"uller}},
  \bibinfo{journal}{Phys.~Rev.~B} \textbf{\bibinfo{volume}{54}},
  \bibinfo{pages}{11169} (\bibinfo{year}{1996}{\natexlab{a}}).

\bibitem[{\citenamefont{Kresse}(1993)}]{Kre93a}
\bibinfo{author}{\bibfnamefont{G.}~\bibnamefont{Kresse}}, Ph.D. thesis,
  \bibinfo{school}{Technische Universit{\"a}t Wien, unpublished}
  (\bibinfo{year}{1993}).

\bibitem[{\citenamefont{Kresse and Furthm\"uller}(1996{\natexlab{b}})}]{Kre96b}
\bibinfo{author}{\bibfnamefont{G.}~\bibnamefont{Kresse}} \bibnamefont{and}
  \bibinfo{author}{\bibfnamefont{J.}~\bibnamefont{Furthm\"uller}},
  \bibinfo{journal}{Comp.~Mat.~Sci.~} \textbf{\bibinfo{volume}{6}},
  \bibinfo{pages}{15} (\bibinfo{year}{1996}{\natexlab{b}}).

\bibitem[{\citenamefont{Hohenberg and Kohn}(1964)}]{Hoh64}
\bibinfo{author}{\bibfnamefont{P.}~\bibnamefont{Hohenberg}} \bibnamefont{and}
  \bibinfo{author}{\bibfnamefont{W.}~\bibnamefont{Kohn}},
  \bibinfo{journal}{Phys.~Rev.~} \textbf{\bibinfo{volume}{136 B}},
  \bibinfo{pages}{864} (\bibinfo{year}{1964}).

\bibitem[{\citenamefont{Kohn and Sham}(1965)}]{Koh65}
\bibinfo{author}{\bibfnamefont{W.}~\bibnamefont{Kohn}} \bibnamefont{and}
  \bibinfo{author}{\bibfnamefont{L.~J.} \bibnamefont{Sham}},
  \bibinfo{journal}{Phys.~Rev.~} \textbf{\bibinfo{volume}{140 A}},
  \bibinfo{pages}{1133} (\bibinfo{year}{1965}).

\bibitem[{\citenamefont{Vanderbilt}(1990)}]{Van90}
\bibinfo{author}{\bibfnamefont{D.}~\bibnamefont{Vanderbilt}},
  \bibinfo{journal}{Phys.~Rev.~B} \textbf{\bibinfo{volume}{41}},
  \bibinfo{pages}{R7892} (\bibinfo{year}{1990}).

\bibitem[{\citenamefont{Kresse and Hafner}(1994)}]{Kre94}
\bibinfo{author}{\bibfnamefont{G.}~\bibnamefont{Kresse}} \bibnamefont{and}
  \bibinfo{author}{\bibfnamefont{J.}~\bibnamefont{Hafner}},
  \bibinfo{journal}{J.~Phys.~Cond.~Matter} \textbf{\bibinfo{volume}{6}},
  \bibinfo{pages}{8245} (\bibinfo{year}{1994}).

\bibitem[{\citenamefont{Perdew et~al.}(1992)\citenamefont{Perdew, Chevary,
  Vosko, Jackson, Pederson, Singh, and Fiolhais}}]{Per92b}
\bibinfo{author}{\bibfnamefont{J.~P.} \bibnamefont{Perdew}},
  \bibinfo{author}{\bibfnamefont{J.~A.} \bibnamefont{Chevary}},
  \bibinfo{author}{\bibfnamefont{S.~H.} \bibnamefont{Vosko}},
  \bibinfo{author}{\bibfnamefont{K.~A.} \bibnamefont{Jackson}},
  \bibinfo{author}{\bibfnamefont{M.~R.} \bibnamefont{Pederson}},
  \bibinfo{author}{\bibfnamefont{D.~J.} \bibnamefont{Singh}}, \bibnamefont{and}
  \bibinfo{author}{\bibfnamefont{C.}~\bibnamefont{Fiolhais}},
  \bibinfo{journal}{Phys.~Rev.~B} \textbf{\bibinfo{volume}{46}},
  \bibinfo{pages}{6671} (\bibinfo{year}{1992}).

\bibitem[{\citenamefont{Perdew and Zunger}(1981)}]{Per81}
\bibinfo{author}{\bibfnamefont{J.~P.} \bibnamefont{Perdew}} \bibnamefont{and}
  \bibinfo{author}{\bibfnamefont{A.}~\bibnamefont{Zunger}},
  \bibinfo{journal}{Phys.~Rev.~B} \textbf{\bibinfo{volume}{23}},
  \bibinfo{pages}{5048} (\bibinfo{year}{1981}).

\bibitem[{\citenamefont{Ceperley and Alder}(1980)}]{Cep80}
\bibinfo{author}{\bibfnamefont{D.~M.} \bibnamefont{Ceperley}} \bibnamefont{and}
  \bibinfo{author}{\bibfnamefont{B.~J.} \bibnamefont{Alder}},
  \bibinfo{journal}{Phys.~Rev.~Lett.~} \textbf{\bibinfo{volume}{45}},
  \bibinfo{pages}{566} (\bibinfo{year}{1980}).

\bibitem[{\citenamefont{Goedecker and Maschke}(1992)}]{Goe92}
\bibinfo{author}{\bibfnamefont{S.}~\bibnamefont{Goedecker}} \bibnamefont{and}
  \bibinfo{author}{\bibfnamefont{K.}~\bibnamefont{Maschke}},
  \bibinfo{journal}{Phys.~Rev.~B} \textbf{\bibinfo{volume}{45}},
  \bibinfo{pages}{1597} (\bibinfo{year}{1992}).

\bibitem[{\citenamefont{Feynman}(1939)}]{Fey39}
\bibinfo{author}{\bibfnamefont{R.~P.} \bibnamefont{Feynman}},
  \bibinfo{journal}{Phys.~Rev.~} \textbf{\bibinfo{volume}{56}},
  \bibinfo{pages}{340} (\bibinfo{year}{1939}).

\bibitem[{\citenamefont{Pulay}(1969)}]{Pul69}
\bibinfo{author}{\bibfnamefont{P.}~\bibnamefont{Pulay}},
  \bibinfo{journal}{Mol.~Phys.~} \textbf{\bibinfo{volume}{17}},
  \bibinfo{pages}{197} (\bibinfo{year}{1969}).

\bibitem[{\citenamefont{Monkhorst and Pack}(1976)}]{Mon76}
\bibinfo{author}{\bibfnamefont{H.~J.} \bibnamefont{Monkhorst}}
  \bibnamefont{and} \bibinfo{author}{\bibfnamefont{J.~D.} \bibnamefont{Pack}},
  \bibinfo{journal}{Phys.~Rev.~B} \textbf{\bibinfo{volume}{13}},
  \bibinfo{pages}{5188} (\bibinfo{year}{1976}).

\bibitem[{\citenamefont{Methfessel and Paxton}(1989)}]{Met89}
\bibinfo{author}{\bibfnamefont{M.}~\bibnamefont{Methfessel}} \bibnamefont{and}
  \bibinfo{author}{\bibfnamefont{A.~T.} \bibnamefont{Paxton}},
  \bibinfo{journal}{Phys.~Rev.~B} \textbf{\bibinfo{volume}{40}},
  \bibinfo{pages}{3616} (\bibinfo{year}{1989}).

\bibitem[{\citenamefont{Nye}(1969)}]{Nye69}
\bibinfo{author}{\bibfnamefont{J.~F.} \bibnamefont{Nye}},
  \emph{\bibinfo{title}{Physical Properties of Crystals}}
  (\bibinfo{publisher}{Oxford University Press, London}, \bibinfo{year}{1969}).

\bibitem[{\citenamefont{Chaikin and Lubensky}(1995)}]{Cha95}
\bibinfo{author}{\bibfnamefont{P.~M.} \bibnamefont{Chaikin}} \bibnamefont{and}
  \bibinfo{author}{\bibfnamefont{T.~C.} \bibnamefont{Lubensky}},
  \emph{\bibinfo{title}{Principles of condensed matter physics}}
  (\bibinfo{publisher}{Cambridge University Press, Cambridge},
  \bibinfo{year}{1995}).

\bibitem[{\citenamefont{Landau and Lifshitz}(1970)}]{Lan70}
\bibinfo{author}{\bibfnamefont{L.~D.} \bibnamefont{Landau}} \bibnamefont{and}
  \bibinfo{author}{\bibfnamefont{E.~M.} \bibnamefont{Lifshitz}},
  \emph{\bibinfo{title}{Theory of elasticity}}, vol. \bibinfo{volume}{VII}
  (\bibinfo{publisher}{Pergamon Press, Oxford}, \bibinfo{year}{1970}).

\bibitem[{\citenamefont{Lee et~al.}(1997)\citenamefont{Lee, Jeong, and
  Chang}}]{Lee97}
\bibinfo{author}{\bibfnamefont{I.-H.} \bibnamefont{Lee}},
  \bibinfo{author}{\bibfnamefont{J.-W.} \bibnamefont{Jeong}}, \bibnamefont{and}
  \bibinfo{author}{\bibfnamefont{K.~J.} \bibnamefont{Chang}},
  \bibinfo{journal}{Phys.~Rev.~B} \textbf{\bibinfo{volume}{55}},
  \bibinfo{pages}{5689} (\bibinfo{year}{1997}).

\bibitem[{\citenamefont{Wang et~al.}(1993)\citenamefont{Wang, Yip, Phillpot,
  and Wolf}}]{Wan93}
\bibinfo{author}{\bibfnamefont{J.}~\bibnamefont{Wang}},
  \bibinfo{author}{\bibfnamefont{S.}~\bibnamefont{Yip}},
  \bibinfo{author}{\bibfnamefont{S.~R.} \bibnamefont{Phillpot}},
  \bibnamefont{and} \bibinfo{author}{\bibfnamefont{D.}~\bibnamefont{Wolf}},
  \bibinfo{journal}{Phys.~Rev.~Lett} \textbf{\bibinfo{volume}{71}},
  \bibinfo{pages}{4182} (\bibinfo{year}{1993}).

\bibitem[{\citenamefont{Wang et~al.}(1995)\citenamefont{Wang, Li, Yip,
  Phillpot, and Wolf}}]{Wan95}
\bibinfo{author}{\bibfnamefont{J.}~\bibnamefont{Wang}},
  \bibinfo{author}{\bibfnamefont{J.}~\bibnamefont{Li}},
  \bibinfo{author}{\bibfnamefont{S.}~\bibnamefont{Yip}},
  \bibinfo{author}{\bibfnamefont{S.}~\bibnamefont{Phillpot}}, \bibnamefont{and}
  \bibinfo{author}{\bibfnamefont{D.}~\bibnamefont{Wolf}},
  \bibinfo{journal}{Phys.~Rev.~B} \textbf{\bibinfo{volume}{52}},
  \bibinfo{pages}{12627} (\bibinfo{year}{1995}).

\bibitem[{\citenamefont{Libotte and Gaspard}(2000)}]{Lib00}
\bibinfo{author}{\bibfnamefont{H.}~\bibnamefont{Libotte}} \bibnamefont{and}
  \bibinfo{author}{\bibfnamefont{J.-P.} \bibnamefont{Gaspard}},
  \bibinfo{journal}{Phys.~Rev.~B} \textbf{\bibinfo{volume}{62}},
  \bibinfo{pages}{7110} (\bibinfo{year}{2000}).

\bibitem[{\citenamefont{Ga\'al-Nagy}(2004)}]{Gaa04d}
\bibinfo{author}{\bibfnamefont{K.}~\bibnamefont{Ga\'al-Nagy}}, Ph.D. thesis,
  \bibinfo{school}{Universit{\"a}t Regensburg,
  http://www.opus-bayern.de/uni-regensburg/volltexte/2004/400}
  (\bibinfo{year}{2004}).

\end{thebibliography}


\end{document}